\documentclass[5p,10pt]{elsarticle}

\biboptions{sort&compress}

\usepackage{amsmath}
\usepackage{geometry}
\usepackage{graphicx}
\usepackage[colorlinks=true,linkcolor=magenta,citecolor=cyan]{hyperref}
\usepackage{booktabs}
\usepackage{tabularx}
\usepackage{placeins}

\journal{Current Opinion in Structural Biology}

\begin{document}
\begin{frontmatter}

\title{Computational design of intrinsically disordered proteins}

\author[1,2]{Giulio Tesei\fnref{fn1}}
\author[1]{Francesco Pesce\fnref{fn1}}
\author[1]{Kresten Lindorff-Larsen\corref{cor1}}
\ead{lindorff@bio.ku.dk}

\cortext[cor1]{Corresponding author}
\fntext[fn1]{These authors contributed equally to this work.}

\affiliation[1]{organization={Structural Biology and NMR Laboratory \& the Linderstr{\o}m-Lang Centre for Protein Science, Department of Biology, University of Copenhagen},
addressline={Ole Maal{\o}es Vej 5},
postcode={2200},
postcodesep={},
city={Copenhagen},
country={Denmark}}
\affiliation[2]{organization={Department of Biomedical Science, Faculty of Health and Society, Malm{\"o} University},
postcode={20506},
postcodesep={},
city={Malm{\"o}},
country={Sweden}}

\begin{abstract}
Protein design has the potential to revolutionize biotechnology and medicine. While most efforts have focused on proteins with well-defined structures, increased recognition of the functional significance of intrinsically disordered regions, together with improvements in their modeling, has paved the way to their computational de novo design. This review summarizes recent advances in engineering intrinsically disordered regions with tailored conformational ensembles, molecular recognition, and phase behavior. We discuss challenges in combining models with predictive accuracy with scalable design workflows and outline emerging strategies that integrate knowledge-based, physics-based, and machine-learning approaches. 
\end{abstract}

\end{frontmatter}

\section*{Introduction}

\begin{figure}[ht!]
\centering
\includegraphics[width=\linewidth]{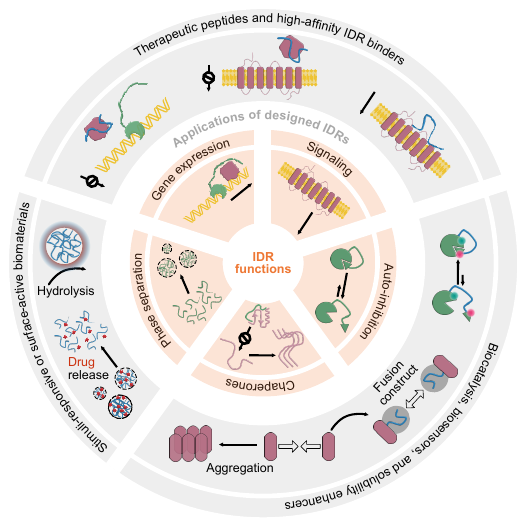}
\caption{Technological applications of IDRs inspired by biological functions. IDRs and flexible peptides can bind to membrane receptors to trigger downstream signaling cascades or act as scaffolds that bring together other signaling molecules \citep{Holehouse2024}. Therefore engineered IDRs have great therapeutic potential. Structured proteins can be designed as high-affinity binders that selectively target IDRs or peptides and may be used to modulate their activity in cell signaling or in regulation of gene expression \citep{wu2025design}.
The conformational properties of IDRs can be harnessed to fine-tune the activity of enzymes or biosensors \citep{Emenecker2023}, while their chaperone functions can be exploited to design fusion proteins with enhanced solubility and stability in formulations \citep{garg2024design}.
In addition, IDRs can form biomolecular condensates that may be designed as stimuli-responsive biomaterials, e.g., for drug delivery \citep{garg2024design}, or to modulate enzyme functions \citep{andre2025toward}. The interfaces of such condensates can also be engineered, for example to enhance or suppress catalytic activity \citep{dai2023interface,Schneider2025,Chen2025}.}
\label{fig:1}
\end{figure}

Around 70 \% of human proteins have at least one stretch of 30 or more amino acids lacking stable secondary or tertiary structure, with approximately 5 \% of human proteins being fully disordered \citep{Holehouse2024,TeseiTrolle2024}. These intrinsically disordered protein regions (IDRs) are highly dynamic and their conformational properties need to be described by large ensembles of rapidly interchanging conformations \citep{Holehouse2024}. These conformational properties are in turn sensitive to solution conditions, post-translational modifications, molecular crowding, the presence of covalently bound folded domains, and interactions with other molecules. Due to their structural heterogeneity and the ability to respond to external cues, an IDR can often perform multiple context-dependent functions \citep{Holehouse2024}. Inspired by their versatility, increasing efforts have focused on exploring strategies for engineering IDRs for technological applications \citep{garg2024design}. In this review, we provide a brief overview of recent advances in de novo design of IDRs focusing on computational approaches (Fig.~\ref{fig:1}).

IDRs are found in various structural contexts: as loops with both ends tethered to the same folded domain, as linkers connecting two domains, or as terminal regions attached to a single domain at one end. Linkers allow the domains they tether to explore distributions of distances and orientations. This flexibility can result in allosteric effects, where a perturbation at one site triggers conformational or dynamical changes at distant sites \citep{garg2024design}. For example, in enzymes involved in cell signaling, linkers mediate the interactions between the domains they connect and influence the equilibrium between active and inactive states, since proximity between the domains often results in auto-inhibition \citep{garg2024design}. Post-translational modifications can perturb the conformational ensemble of the linker, enabling the enzyme to respond to external signals by switching state. Moreover, the volume explored by the conformational ensemble of a linker determines the local effective concentration of the tethered domain, thereby influencing binding reactions \citep{garg2024design}. Terminal IDRs can instead act as entropic bristles that enhance solubility and sterically prevent protein aggregation by increasing the excluded volume \citep{Holehouse2024}.

Characterizing the conformational ensembles of IDRs is a complex task that often requires combinations of multiple experimental techniques and computational modeling \citep{ghafouri2025towards}. While for globular proteins the first step in de novo design is often to identify amino acid sequences that stably fold into specific structures \citep{Zhang2025cosb}, the design target for IDRs is a large and heterogeneous ensemble of conformations. In the first part of this review, we discuss how computational methods are contributing to the design of IDRs with bespoke conformational properties, which is an area of growing interest in biotechnology \citep{garg2024design}.

A large conformational ensemble enables the same IDR to interact with multiple partners through distinct binding modes, and the biological function of IDRs is tightly coupled to molecular recognition through interactions with other biomolecules \citep{Holehouse2024}. How IDRs bind and interact with their partners is an active field of research and different models have been proposed. IDRs that are entirely disordered when free in solution can fold when interacting with a specific partner via a mechanism referred to as folding-upon-binding. Other IDRs bind while maintaining their disordered character through fuzzy binding modes, which can cover a wide range of structural heterogeneity, from a few fixed binding modes to a completely diffuse binding of an IDR on the surface of its partner \citep{Holehouse2024}. As described in the second section of this review, computational methods have advanced significantly in enabling the design of IDR-mediated binding. 

Through multivalent intermolecular interactions, IDRs often contribute to the formation of biomolecular condensates: dynamic assemblies of primarily proteins and nucleic acids that compartmentalize the cellular environment. Unlike membrane-bound organelles, biomolecular condensates are not enclosed by lipid membranes and their formation and material properties are sensitive to biochemical signals and environmental changes. Therefore, biomolecular condensates are involved in the regulation of critical biological processes such as cell signaling and response to stress. 
The material properties of biomolecular condensates are vicoelastic and can vary from predominantly viscous to predominantly elastic \citep{Pappu2023}. This wide range of dynamical properties is modulated by IDR-mediated interactions which give rise to a network of physical cross-links spanning the condensate \citep{Pappu2023}. In the third section of this review, we discuss the design of IDRs forming or partitioning to biomolecular condensates, with tailored properties for use in therapeutic strategies and stimuli-responsive biomaterials.

\section*{Conformational properties and linker function}
Designing IDRs with targeted functions requires knowledge of how the amino acid sequence of an IDR is related to protein function through the conformational ensemble that it encodes, both in the context of the isolated protein and of the full-length multi-domain protein. Amino acid composition and patterning have been related to IDR compaction as well as to specific biological functions and selective cellular compartmentalization. Most IDRs are polyampholytes \citep{Das2013}, for which both the fraction of charged residues and their segregation into blocks of opposite charge are conserved features \cite{Zarin2019,TeseiTrolle2024} that determine compaction and are key for biological functions \citep{Holehouse2024}. The patterning of aromatic residues has also been identified as a conserved sequence feature. 
For example, in prion-like domains, aromatics are significantly more uniformly spaced along the linear sequence than expected from random distributions \cite{Martin2020,Cohan2022}. The investigation of sequence-ensemble-function relationships of IDRs has benefited from the development of physics-inspired sequence features quantifying linear patterning, such as $\kappa$ \citep{Das2013}, sequence charge decoration, sequence hydropathy decoration \citep{Zheng2020}, and additional sets of sequence features that capture other aspects of IDR sequence composition and patterning \citep{Zarin2019,Cohan2022}. Proteome-wide analyses can identify features associated with specific cellular functions or localizations \citep{pritivsanac2024functional,Ruff2025}, which can then be rationally introduced in the design of new sequences \citep{Strome2023,Ruff2025}.

\begin{figure*}[ht!]
\centering
\includegraphics[width=\linewidth]{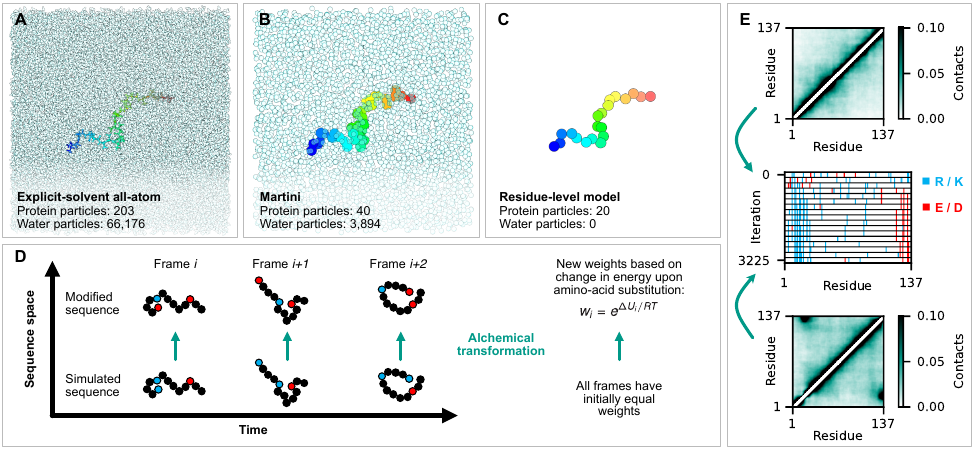}
\caption{Simulation-based approach for designing IDRs with bespoke conformational properties. Top panels illustrate molecular force fields with different resolutions for an alanine 20-mer: (A) all-atom model with explicit solvent, where each atom is a particle; (B) coarse-grained Martini model, where each alanine residue is described by two beads (backbone and side-chain) and each water bead represents four water molecules; (C) residue-level model with one bead per residue and implicit water. (D) Alchemical transformations can be applied to the simulated conformational ensemble of an IDR by modifying the chemical identity of one or more particles, generating a new sequence. Statistical weights can be computed for each frame by comparing potential energies between the modified and original sequences. These weights allow conformational properties of the modified sequence to be estimated as a weighted average over the ensemble of the original sequence, for example using the multistate Bennett acceptance ratio \citep{Pesce2024}. (E) Example of alchemical transformations applied to design a sequence matching a target contact map, characterized by frequent contacts between N- and C-terminal residues \citep{Pesce2024}. Starting from an initial sequence lacking favorable interactions between the termini, iteratively swapping the positions of randomly selected residue pairs results in a sequence with enhanced charge segregation (middle) which closely matches the target contact map (bottom).}
\label{fig:2}
\end{figure*}

Molecular simulations have been instrumental in furthering our understanding of sequence-ensemble relationships and enabling the design of IDRs. All-atom simulations with explicit water are computationally expensive, especially for IDRs, due to the diverse sets of structures they may attain and the large number of water molecules required to minimize finite-size effects for extended protein conformations (Fig.~\ref{fig:2}A). More efficient models have been developed by reducing the number of particles. This can be achieved either through implicit descriptions of the solvent or by coarse-graining protein and water molecules. Examples include the implicit-solvent all-atom ABSINTH model \citep{Vitalis2009}, the explicit-solvent coarse-grained Martini model \citep{Souza2021} (Fig.~\ref{fig:2}B), and residue-level implicit-solvent models with amino acid-specific parameters informed by hydrophobicity scales \citep{Dignon2018,Dannenhoffer-Lafage2021,Regy2021,Joseph2021,Tesei2021,Jussupow2025} (Fig.~\ref{fig:2}C). The computational efficiencies of these models has enabled the characterization of conformational ensembles across entire proteomes \citep{TeseiTrolle2024,Lotthammer2024}, shedding light on how sequence features shape the conformational properties of IDRs. These insights, in turn, have informed the rational design of IDRs with targeted functional outcomes. 

Using as model system the intracellular IDR of a membrane receptor regulating signaling pathways involved in cell differentiation and cancer, ABSINTH simulations showed that increasing charge segregation at fixed amino acid composition can lead to increased chain compaction \citep{sherry2017control}. Experiments further revealed that IDR compaction correlates with weaker binding with downstream proteins in the signaling pathway and lower transcriptional activity, consistent with a model in which increased compaction results in a reduction in the effective concentration of the binding sites located in the intracellular domain \citep{sherry2017control}.

Drawing on previous work \citep{Harmon2016,Zeng2021}, recent studies employed molecular simulations to systematically design IDRs through automated procedures \citep{Pesce2024,Krueger2024}. In one study, we combined the residue-level CALVADOS model with Monte Carlo sampling in sequence space to design proteins with increased compaction, while preserving overall amino acid composition \citep{Pesce2024}. We used alchemical calculations to perform efficient trajectory reweighting to evaluate the effect of sequence changes on the conformational ensemble (Fig.~\ref{fig:2}D). In this approach, energy differences between initial and trial sequences were computed across all sampled conformations, allowing statistical weights to be assigned to each structure. These weights were then used to estimate ensemble-averaged observables by reweighting the original simulation trajectory, with additional simulations required only when cumulative sequence changes made the estimated weights unreliable. Experimental validation of selected design variants indicated that the procedure provided accurate predictions for sequences differing considerably in charge patterning, as well as from the natural sequences used to parameterize the CALVADOS model. The simulation-driven design can be computationally expensive; as a complement we showed that it is also possible to design sequences using a fast machine-learning model \citep{Pesce2024} that was previously trained on CALVADOS simulations to predict chain compaction directly from sequence \citep{TeseiTrolle2024}. Finally, as a more complex design target than chain compaction, we also showed that it is possible to design sequences with a specific intramolecular contact map (Fig.~\ref{fig:2}E). 

A related framework combined reweighting of simulations using the residue-level Mpipi-GG model with gradient-based optimization \citep{Krueger2024}. This approach enabled the efficient design of IDRs with a broader range of targeted ensemble properties, including sequences with compact or expanded conformations, decoupled relationships between radius of gyration and end-to-end distance, responsiveness to changes in salt concentration, and strong binding affinity to target IDRs.

Together, these types of studies demonstrate how physics-based models can guide sequence optimization for IDR design. Tuning charge patterning emerged as an effective strategy to control chain compaction and environmental responsiveness, as further supported by experiments in live cells under osmotic perturbation using synthetic IDRs designed with the GOOSE software package \citep{Emenecker2023}.

Machine-learning models are emerging as accurate and computationally efficient tools for designing IDRs with targeted conformational properties. Several have been trained on large datasets derived from coarse-grained simulations of isolated IDRs, enabling them to learn sequence-ensemble relationships. These include models that predict ensemble-averaged conformational properties \citep{Zheng2020,TeseiTrolle2024,Lotthammer2024,Mollaei2024} and diffusion-based generative models that reconstruct full conformational ensembles from sequence \citep{janson2023direct,Novak2025}. These tools, used individually or in combination, enable efficient design of IDRs with targeted conformational features, such as intrachain distance maps \citep{Novak2025}. However, their performance depends on the physics-based model used to generate the training data, and common limitations include lack of secondary structure information. 

Other generative models were trained on atomistic-resolution ensembles alone \citep{Janson2024} or in combination with other sources, such as CALVADOS simulations, all-atom force fields, Protein Data Bank or AlphaFold structures \citep{Zhu2024,zhang2025,Schnapka2025}, and experimental protein stability measurements \citep{Lewis2025}. These models predict IDR ensembles with secondary structure propensities, either for IDR in isolation \citep{Janson2024,Zhu2024} or in the context of the full-length protein \citep{zhang2025,Lewis2025,Schnapka2025}, and have the potential for being used in sequence design.

In contrast to structure-based models, sequence-based generative models enable de novo design of IDRs without the need for protein structures in their training sets and instead rely on evolutionary information, including from multiple sequence alignments of orthologous sequences \citep{Alamdari2024}. While significantly more efficient than molecular simulations, these models have only recently begun to account for environmental conditions \citep{Janson2025temp}, and thus their applicability for designing sequences that respond to environmental stimuli has been limited. Moreover, because they learn patterns in the training data rather than the physics underlying the folding landscape, they may struggle to design multi-domain proteins with folds or domain architectures that differ significantly from those in nature.

Interactions between IDRs and the structured domains they tether can significantly influence the conformational ensemble of the IDR and are often critical for function \citep{mittal2018sequence,taneja2021folded}. Fast models of full-length multi-domain proteins---such as BioEmu \citep{Lewis2025}, IDPForge \citep{zhang2025}, EvoDiff \citep{Alamdari2024}, CALVADOS~3, and AFflecto \citep{pajkos2025afflecto}---are therefore promising tools for IDR design. However, recent benchmarking efforts, including CASP16 \citep{mcbride2025predicting}, point to ongoing challenges in accurately predicting the conformational ensembles of non-globular proteins containing IDRs.

\section*{Binding and molecular recognition}
Protein design strategies for IDR binding have addressed both the generation of IDR sequences with targeted secondary-structure propensities in the bound state \citep{Harmon2016} and the design of protein binders with high affinity for IDRs in structured or extended conformations \citep{wu2025design,liu2025diffusing,Bhat2025,Chen2025PepMLM}. 

IDRs that form helices in the bound state often display intrinsic helicity in the unbound form, which may contribute to the thermodynamic and kinetic stability of binding. To probe this relationship, a physics-based computational approach, combining ABSINTH simulations with a genetic algorithm, was developed to design IDRs matching an input helicity profile under the constraint of fixed amino acid composition and fixed positions of residues critical for complex formation with the folded partner \citep{Harmon2016}. In this context, the genetic algorithm takes inspiration from the process of natural selection, iteratively modifying a sequence and evaluating its fitness based on how closely it matches the target helicity. High-fitness sequences are selected to generate offspring through further modifications, and this process is repeated over successive generations until a sequence that meets the design objective is identified.

For the complementary task of designing folded domains that bind to an IDR, knowledge-based design methods such as Rosetta, along with generative diffusion models such as RFdiffusion, have been used to engineer pockets and grooves in folded proteins that accommodate peptides in polyproline II, $\alpha$-helical, or $\beta$-strand conformations. These methods achieve high affinity by satisfying steric constraints and compensating the loss in solvation and conformational entropy with energetically favorable interactions \citep{vazquez2024novo,sahtoe2024design}. For this task, RFdiffusion was developed using structures of two-chain complexes from the Protein Data Bank. Noise was progressively added to the coordinates of one chain at a time, and the model was trained to recover the initial conformation of one chain in the context of the fixed structure of the second chain \citep{vazquez2024novo}.

However, in many cases, the bound conformation of the IDR is not known, and many IDRs form fuzzy complexes where they adopt multiple conformations. To address this challenge, Rosetta and RFdiffusion were recently combined to design high-affinity binders for IDRs without predefined structural preferences \citep{wu2025design,liu2025diffusing}. The general approach relies on the ability of the binder to guide subregions of the IDR to adopt bound conformations which can be accommodated in its pocket.
In one approach, Rosetta methods are used to generate a library of template structures by assembling scaffolds with pockets specialized to target specific dipeptides. A fragment of the IDR, selected to minimize off-target binding, is threaded through the templates and the backbone conformation and sequence of the binder in these complexes is further refined using deep-learning models, including AlphaFold2, ProteinMPNN, and RFdiffusion \citep{wu2025design}. In a different approach, RFdiffusion is given only the IDR sequence as input and is used to generate backbone structures for the entire complex by sampling multiple conformations of both IDR and binder simultaneously, thus allowing one to adapt to the other \citep{liu2025diffusing}. The binders with highest affinity generated by this method induce parts of the IDR to adopt $\alpha$-helical or $\beta$-strand conformation in the bound state \citep{liu2025diffusing} whereas the complementary method based on threading through scaffold templates generates high-affinity binding for targets in more extended conformations \citep{wu2025design}.

To complement these primarily structure-based approaches, a pipeline based entirely on sequence-based models has been developed to generate short linear peptides that bind with high affinity to diverse protein targets \citep{Bhat2025,Chen2025PepMLM}. Using protein language models trained on sequences of thousands of peptide–protein complexes, this approach enables the design of peptides that bind with specificity to target proteins with potentially unknown structural features. As an example of potential applications, designed peptides fused to an E3 ubiquitin ligase conferred selective degradation of pathogenic proteins in cells \citep{Bhat2025,Chen2025PepMLM}.

In addition to sequence-specific binding to structured interfaces, IDRs can also form high-affinity complexes through chemically specific interactions in which the two bound proteins are highly dynamic. These binding reactions often involve charge complementarity or multivalent interactions between charged or aromatic residues \citep{Holehouse2024}. Recently, a sequence-based method was developed to predicts two-chain interaction maps using amino acid–specific parameters derived from residue-level force fields, providing an efficient tool for rational design of binders forming fully disordered complexes \citep{ginell2025sequence}.

\begin{figure*}[tbp]
\centering
\includegraphics[width=\linewidth]{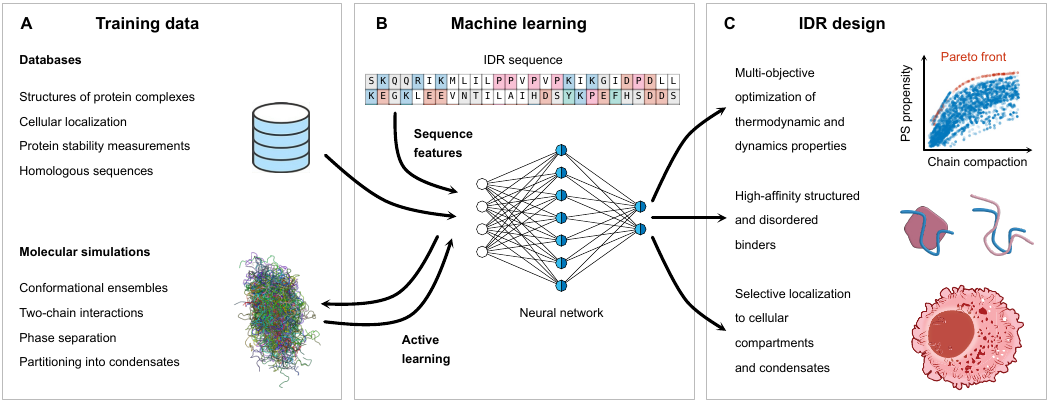} \caption{Role of machine learning in de novo design of IDRs. (A)~Machine-learning models can be trained on diverse data sources, from molecular dynamics simulations to annotations of cellular localization and protein structures from the Protein Data Bank. (B)~Often implemented as neural networks using sequence-encoded features as input, these models can initially be trained on a limited region of sequence space as surrogate models. Through active learning, additional simulations are performed during the design campaign to generate new data, and the surrogate model is retrained on the expanded dataset to progressively improve its accuracy. (C)~Machine-learning models have been developed to predict biophysical observables, biological annotations, and protein structures. When combined, machine-learning models can be used to identify a set of sequences that strike a trade-off between multiple design objectives, defining a Pareto front.}
\label{fig:3}
\end{figure*}

\section*{Stimuli-responsive phase separation and selective partitioning}
IDRs can serve as scaffold proteins that drive the formation of biomolecular condensates. These microenvironments with distinct and emergent physicochemical properties are highly permeable and can selectively recruit client molecules from the surrounding medium. Since phase separation (PS) is sensitive to environmental conditions, biomolecular condensates can sequester and release clients in response to external stimuli. These functions can be engineered for biocatalysis and drug delivery \citep{dai2023programmable,gil2024local,liang2024situ}.

Early work on engineering IDRs driving PS focused on elastin-like polypeptides (ELPs), synthetic IDRs inspired by the sequence and properties of tropoelastin, the precursor of elastin fibers in the extracellular matrix \citep{urry1974synthetic}. ELPs were first designed as tandem repeats of a hydrophobic pentapeptide \citep{urry1974synthetic} which phase separate above a lower critical solution temperature (LCST). Varying chain length and the identity of a single amino acid in the pentapeptide, these synthetic IDRs served as a model system for understanding and engineering the relationship between sequence features and phase behavior \citep{urry1992hydrophobicity,mcdaniel2013unified}. The strategy of scanning amino acid composition in synthetic IDRs based on minimal repeat units was extended to resilin-like polypeptides, polyampholytes that phase separate below an upper critical solution temperature, leading to a set of rules linking composition to phase behavior \citep{quiroz2015sequence,dzuricky2020novo}. Computational methods have further expanded the sequence space accessible to ELP design \citep{Zeng2021}. Leveraging coil-to-globule transitions of single chains as a proxy for the temperature dependence of PS, ABSINTH simulations were combined with a genetic algorithm to identify novel candidate pentapeptides predicted to display LCST behavior \citep{Zeng2021}.

As the role of biomolecular condensates in biological processes has become increasingly appreciated, research on the sequence dependence of PS has focused on how amino acid substitutions affect homotypic PS in naturally occurring IDRs. 
The systematic quantification of the saturation concentration, $c_\textit{sat}$---the protein concentration in the dilute phase at equilibrium with the condensate---has led to the distinction between residues that promote PS more strongly and residues that modulate PS primarily through their solvation volume \citep{Bremer2022}. Moreover, the approach has enabled the ranking of amino acid-specific contributions to PS propensity \citep{Bremer2022,Norrild2024}. 

This body of work on the sequence dependence of PS has also informed the development of coarse-grained models of IDRs \citep{Martin2020,Regy2021,Joseph2021,Tesei2021,Farag2022,Jussupow2025}. Capturing experimental PS propensities, these models have served as the basis for automated sequence explorations aimed at optimizing target thermodynamic and dynamic properties of biomolecular condensates. For example, genetic algorithms have been employed to evolve sequences toward increased PS propensities \citep{lichtinger2021targeted} or to form multiphasic condensates, i.e., two-component systems with preferential partitioning of one component in the core or at the surface of the condensate formed by the other component \citep{chew2023thermodynamic}. These studies used fitness functions formulated as linear combinations of the concentrations in the dense and dilute phases and provided insights into how the relative strengths of homotypic and heterotypic interactions give rise to multiphasic condensates \citep{chew2023thermodynamic}. 

Converging $c_\textit{sat}$ predictions from molecular simulations can be computationally demanding. In design frameworks where many sequences need to be evaluated, one strategy to circumvent these simulations exploits the coupling between conformational and phase propensity. As discussed in the first section of this review, sequences designed to maximize single-chain compaction, under the constraint of constant amino acid composition, generally exhibit enhanced PS propensity \cite{Pesce2024}. Similarly, PS propensity can be estimated from contact maps or second virial coefficients calculated from two-chain simulations \citep{jin2025predicting}. These metrics, however, have limitations due to the different relative effects of short- and long-range interactions in determining chain compaction, two-chain self-association, and multi-chain PS \citep{oliver2025b2enoughevaluatingsimple}.

Moving beyond global interaction metrics to residue-level descriptions, coarse-grained force field parameters have been used to predict intermolecular interaction maps directly from sequence. This approach bypasses the need to perform simulations and allows for rapid proteome-scale predictions of homo- and heterotypic interactions, which can be related to PS propensities \citep{ginell2025sequence}. As an alternative, machine learning models have been developed by training on PS data from molecular simulations (Fig.~\ref{fig:3}). Generating a sufficiently large training set spanning a wide range of PS propensities is computationally expensive, prompting the use of active learning strategies that iteratively enrich the training set with sequences where the model is least confident \citep{vonBulow2025,An2024} (Fig.~\ref{fig:3}A,B). Using a small set of sequence-derived parameters as input features, small neural networks were trained on CALVADOS simulation data to predict $c_\textit{sat}$ and the free energy of transfer between dilute and dense phases \citep{vonBulow2025}. These models enable sequence exploration and design on a large scale and have been used to generate IDRs with increased PS propensity under various constraints, revealing that, while compaction correlates with phase behavior at constant composition, this relationship may break down when net charge and hydropathy are allowed to vary \citep{vonBulow2025}. 

In workflows for de novo design of IDRs, machine-learning models may also serve as surrogate models that are progressively refined as more data are generated during the design campaign \citep{An2024,Changiarath2025} (Fig.~\ref{fig:3}A,B). Active learning and Bayesian optimization with uncertainty quantification have been used to efficiently explore sequence space by balancing exploitation of promising candidates (predictions closer to the target) and exploration of under-sampled regions (higher uncertainty of the predictions) \citep{An2024,Changiarath2025}. This approach enabled the design of sequences that achieve a trade-off between condensate stability and internal dynamics \citep{An2024}, as well as the design of client IDRs partitioning at the surface or in the core of a condensate formed by a scaffold protein \citep{Changiarath2025}. 

The selective partitioning and enrichment of proteins into condensates and other subcellular compartments is a key factor for the efficacy of protein therapeutics. 
Sequence-based approaches have been shown to enable the design of IDRs that concentrate in specific subcellular compartments. One strategy uses statistical analyses to identify sequence features enriched in IDRs known to localize to compartments due to their distinct chemical environments. Based on these analyses, iterative amino acid substitutions were used to generate synthetic sequences that match the identified features, driving selective localization \citep{Strome2023}. Alternatively, a deep-learning model was trained on curated annotations of protein compartmentalization to predict selective partitioning into various biomolecular condensates \citep{Kilgore2025}. These approaches point to a promising research direction while also highlighting the challenges of predicting subcellular partitioning from general sequence features, especially in cases where localization cannot be inferred from evolutionary relatedness to known proteins.

Due to asymmetric distributions of salt ions and water molecules between the dense and dilute phases, biomolecular condensates formed by IDRs were shown to exhibit interphase electric potentials and pH gradients \citep{Ausserwoeger2024,posey2024biomolecular}. As amino acids, salt ions, and water compositions change across the boundary between the two phases, the interface is characterized by distinct physicochemical properties, which can catalyze chemical reactions \citep{dai2023interface,Chen2025} and facilitate the nucleation of amyloid fibrils in neurodegenerative diseases \citep{Linsenmeier2023,das2025tunable}. The selective partitioning of peptides or client IDRs at the interface of biomolecular condensates can effectively modulate such properties, offering a promising strategy for engineering catalytic functions and developing therapeutic interventions \citep{Chen2025,visser2024controlling,Schneider2025}. De novo design of peptides that localize to the interface of condensates was achieved with an optimization pipeline integrating molecular simulations and surrogate models refined via active learning \citep{Schneider2025}. The pipeline identifies peptides that strike a trade-off between minimal self-association and maximal partitioning to the interface. Optimal sequences were designed and experimentally validated for condensates formed by three distinct scaffold IDRs. In all cases, the peptides exhibited amphiphilic character, promoting entry while preventing full partitioning \citep{Schneider2025}. 
The combination of molecular simulations and neural networks is also promising for designing IDRs active at the water–membrane interface, as previously shown for the design of peptides that selectively bind to curved membranes \citep{vanHilten2023}.

\section*{Future perspectives}
Recent advances in computationally efficient and accurate molecular models of IDRs have enabled exploration of sequence-ensemble relationships at unprecedented scale \citep{TeseiTrolle2024,Lotthammer2024}. Machine learning has further expanded these predictions and, when trained on diverse data sources, extended them to other properties, including structures of bound IDRs within complexes \citep{wu2025design,liu2025diffusing} and cellular localization. These developments now make it possible to design IDRs, with potential applications ranging from engineering enzymes with linkers optimized for activity, to creating high-affinity binders targeting IDRs, and designing IDRs that selectively partition into specific cellular compartments. Combining coarse-grained simulations with machine-learning surrogate models and active learning has emerged as a robust strategy applicable to simultaneously optimize multiple competing properties of systems of higher complexity, such as multicomponent biomolecular condensates \citep{An2024,Changiarath2025,Schneider2025}. Because molecular simulations can incorporate environmental effects, tuning condensate stability, surface activity, and dynamics through sequence offers a route to designing functional, stimuli-responsive biomaterials. Sequence-based machine-learning models can extract information from evolutionary relationships that can be used in design \citep{Alamdari2024,Bhat2025,Chen2025PepMLM}. IDR sequences, however, evolve rapidly and it may be difficult to use methods that explicitly or implicitly rely on sequence alignments \citep{Holehouse2024}. Alignment-free approaches \citep{Zarin2019,Cohan2022,Ruff2025} may provide an alternative representation of IDR sequences that can help guide IDR design \citep{Strome2023}

However, important limitations remain. Current molecular models often neglect transient secondary structure, ion-specific effects, and the temperature dependence of hydrophobic interactions. Moreover, current models for non-globular proteins struggle to accurately reproduce experimental observables \citep{mcbride2025predicting}. Generative models trained on simulations inherit these limitations, whereas deep-learning models trained on heterogeneous datasets may have limited transferability due to overlapping training and test sets or poorly defined independent benchmarks \citep{Janson2025}. Addressing these challenges will improve the accuracy of predictions for conformational ensembles and phase behavior of IDRs, and we envision these developments as crucial for IDR design. While accurately designing conformational ensembles or phase properties of IDRs does not guarantee control over their biological function, computational approaches also facilitate systematic investigations that improve our understanding of the relationships between sequence, biophysical properties, and function, which will ultimately contribute to more effective targeting of desired functions.

\section*{Acknowledgments}

This work is a contribution from the PRISM (Protein Interactions and Stability in Medicine and Genomics) centre funded by the Novo Nordisk Foundation (to K.L.-L.; NNF18OC0033950). G.T. acknowledges support from the Swedish Research Council (2024-04539\_VR).

\section*{Conflicts of Interest}
K.L.-L. holds stock options in, receives sponsored research from, and is a consultant for Peptone. The remaining authors declare no competing interests.

\FloatBarrier

\bibliographystyle{elsarticle-num}

{\footnotesize
\bibliography{references}}

\begin{thebibliography}{10}
\expandafter\ifx\csname url\endcsname\relax
  \def\url#1{\texttt{#1}}\fi
\expandafter\ifx\csname urlprefix\endcsname\relax\def\urlprefix{URL }\fi
\expandafter\ifx\csname href\endcsname\relax
  \def\href#1#2{#2} \def\path#1{#1}\fi

\bibitem{Holehouse2024}
A.~S. Holehouse, B.~B. Kragelund, The molecular basis for cellular function of
  intrinsically disordered protein regions, Nature Reviews. Molecular Cell
  Biology 25~(3) (2024) 187--211.
\newblock \href {https://doi.org/10.1038/s41580-023-00673-0}
  {\path{doi:10.1038/s41580-023-00673-0}}.

\bibitem{TeseiTrolle2024}
G.~Tesei, A.~I. Trolle, N.~Jonsson, J.~Betz, F.~E. Knudsen, F.~Pesce, K.~E.
  Johansson, K.~{Lindorff-Larsen}, Conformational ensembles of the human
  intrinsically disordered proteome, Nature 626~(8000) (2024) 897--904.
\newblock \href {https://doi.org/10.1038/s41586-023-07004-5}
  {\path{doi:10.1038/s41586-023-07004-5}}.

\bibitem{garg2024design}
A.~Garg, N.~S. Gonz{\'a}lez-Foutel, M.~B. Gielnik, M.~Kjaergaard, Design of
  functional intrinsically disordered proteins, Protein Engineering, Design and
  Selection 37 (2024) gzae004.
\newblock \href {https://doi.org/10.1093/protein/gzae004}
  {\path{doi:10.1093/protein/gzae004}}.

\bibitem{ghafouri2025towards}
H.~Ghafouri, P.~Kadeřávek, A.~M. Melo, M.~C. Aspromonte, P.~Bernadó,
  J.~Cortes, Z.~Dosztányi, G.~Erdos, M.~Feig, G.~Janson, K.~Lindorff-Larsen,
  F.~A.~A. Mulder, P.~Nagy, R.~Pestell, D.~Piovesan, M.~Schiavina, B.~Schuler,
  N.~Sibille, G.~Tesei, P.~Tompa, M.~Vendruscolo, J.~Vondrasek, W.~Vranken,
  L.~Zidek, S.~C.~E. Tosatto, A.~M. Monzon, Towards a unified framework for
  determining conformational ensembles of disordered proteins (2025).
\newblock \href {https://doi.org/10.48550/arXiv.2504.03590}
  {\path{doi:10.48550/arXiv.2504.03590}}.

\bibitem{Zhang2025cosb}
Z.~Zhang, C.~Ou, Y.~Cho, Y.~Akiyama, S.~Ovchinnikov, Artificial intelligence
  methods for protein folding and design, Current Opinion in Structural Biology
  93 (2025) 103066.
\newblock \href {https://doi.org/10.1016/j.sbi.2025.103066}
  {\path{doi:10.1016/j.sbi.2025.103066}}.

\bibitem{Pappu2023}
R.~V. Pappu, S.~R. Cohen, F.~Dar, M.~Farag, M.~Kar, Phase {{Transitions}} of
  {{Associative Biomacromolecules}}, Chemical Reviews 123~(14) (2023)
  8945--8987.
\newblock \href {https://doi.org/10.1021/acs.chemrev.2c00814}
  {\path{doi:10.1021/acs.chemrev.2c00814}}.

\bibitem{wu2025design}
K.~Wu, H.~Jiang, D.~R. Hicks, C.~Liu, E.~Muratspahi{\'c}, T.~A. Ramelot,
  Y.~Liu, K.~McNally, S.~Kenny, A.~Mihut, et~al., Design of intrinsically
  disordered region binding proteins, Science 389~(6757) (2025) eadr8063.
\newblock \href {https://doi.org/10.1126/science.adr8063}
  {\path{doi:10.1126/science.adr8063}}.

\bibitem{Emenecker2023}
R.~J. Emenecker, K.~Guadalupe, N.~M. Shamoon, S.~Sukenik, A.~S. Holehouse,
  Sequence-ensemble-function relationships for disordered proteins in live
  cells, bioRxiv (Jan. 2023).
\newblock \href {https://doi.org/10.1101/2023.10.29.564547}
  {\path{doi:10.1101/2023.10.29.564547}}.

\bibitem{andre2025toward}
A.~A. Andr{\'e}, N.~Rehnberg, A.~Garg, M.~Kj{\ae}rgaard, Toward design
  principles for biomolecular condensates for metabolic pathways, Advanced
  Biology 9~(5) (2025) 2400672.
\newblock \href {https://doi.org/10.1002/adbi.202400672}
  {\path{doi:10.1002/adbi.202400672}}.

\bibitem{dai2023interface}
Y.~Dai, C.~F. Chamberlayne, M.~S. Messina, C.~J. Chang, R.~N. Zare, L.~You,
  A.~Chilkoti, Interface of biomolecular condensates modulates redox reactions,
  Chem (2023).
\newblock \href {https://doi.org/10.1016/j.chempr.2023.04.001}
  {\path{doi:10.1016/j.chempr.2023.04.001}}.

\bibitem{Schneider2025}
T.~N. Schneider, M.~Gil-Garcia, M.~A. B{\"u}hler, L.~F. Santos, L.~Faltova,
  G.~Guill{\'e}n-Gos{\'a}lbez, P.~Arosio, De novo design of peptides localizing
  at the interface of biomolecular condensates, bioRxiv (2025).
\newblock \href {https://doi.org/10.1101/2025.05.09.653111}
  {\path{doi:10.1101/2025.05.09.653111}}.

\bibitem{Chen2025}
M.~W. Chen, X.~Guo, M.~Farag, N.~Qian, X.~Song, A.~Ni, V.~Liu, X.~Yu, Y.~Ma,
  L.~Yang, W.~Yu, M.~R. King, J.~Lee, R.~N. Zare, W.~Min, R.~V. Pappu, Y.~Dai,
  Condenzymes: Biomolecular condensates with inherent catalytic activities,
  bioRxiv (2025).
\newblock \href {https://doi.org/10.1101/2024.07.06.602359}
  {\path{doi:10.1101/2024.07.06.602359}}.

\bibitem{Das2013}
R.~K. Das, R.~V. Pappu, Conformations of intrinsically disordered proteins are
  influenced by linear sequence distributions of oppositely charged residues,
  Proceedings of the National Academy of Sciences 110~(33) (2013) 13392--13397.
\newblock \href {https://doi.org/10.1073/pnas.1304749110}
  {\path{doi:10.1073/pnas.1304749110}}.

\bibitem{Zarin2019}
T.~Zarin, B.~Strome, A.~N. Nguyen~Ba, S.~Alberti, J.~D. {Forman-Kay}, A.~M.
  Moses, Proteome-wide signatures of function in highly diverged intrinsically
  disordered regions, eLife 8 (2019) e46883.
\newblock \href {https://doi.org/10.7554/eLife.46883}
  {\path{doi:10.7554/eLife.46883}}.

\bibitem{Martin2020}
E.~W. Martin, A.~S. Holehouse, I.~Peran, M.~Farag, J.~J. Incicco, A.~Bremer,
  C.~R. Grace, A.~Soranno, R.~V. Pappu, T.~Mittag, Valence and patterning of
  aromatic residues determine the phase behavior of prion-like domains, Science
  367~(6478) (2020) 694--699.
\newblock \href {https://doi.org/10.1126/science.aaw8653}
  {\path{doi:10.1126/science.aaw8653}}.

\bibitem{Cohan2022}
M.~C. Cohan, M.~K. Shinn, J.~M. Lalmansingh, R.~V. Pappu, Uncovering non-random
  binary patterns within sequences of intrinsically disordered proteins,
  Journal of Molecular Biology 434~(2) (2022) 167373.
\newblock \href {https://doi.org/10.1016/j.jmb.2021.167373}
  {\path{doi:10.1016/j.jmb.2021.167373}}.

\bibitem{Zheng2020}
W.~Zheng, G.~Dignon, M.~Brown, Y.~C. Kim, J.~Mittal, Hydropathy {{Patterning
  Complements Charge Patterning}} to {{Describe Conformational Preferences}} of
  {{Disordered Proteins}}, Journal of Physical Chemistry Letters 11~(9) (2020)
  3408--3415.
\newblock \href {https://doi.org/10.1021/acs.jpclett.0c00288}
  {\path{doi:10.1021/acs.jpclett.0c00288}}.

\bibitem{pritivsanac2024functional}
I.~Priti{\v s}anac, T.~R. Alderson, {\DJ}.~Kolari{\'c}, T.~Zarin, S.~Xie,
  A.~Lu, A.~Alam, A.~Maqsood, J.-Y. Youn, J.~D. Forman-Kay, A.~M. Moses, A
  functional map of the human intrinsically disordered proteome, bioRxiv
  (2024).
\newblock \href {https://doi.org/10.1101/2024.03.15.585291}
  {\path{doi:10.1101/2024.03.15.585291}}.

\bibitem{Ruff2025}
K.~M. Ruff, M.~R. King, A.~W. Ying, V.~Liu, A.~Pant, W.~E. Lieberman, M.~K.
  Shinn, X.~Su, C.~Kadoch, R.~V. Pappu, Molecular grammars of intrinsically
  disordered regions that span the human proteome, bioRxiv (2025).
\newblock \href {https://doi.org/10.1101/2025.02.27.640591}
  {\path{doi:10.1101/2025.02.27.640591}}.

\bibitem{Strome2023}
B.~Strome, K.~Elemam, I.~Pritisanac, J.~D. Forman-Kay, A.~M. Moses,
  Computational design of intrinsically disordered protein regions by matching
  bulk molecular properties, bioRxiv (2023).
\newblock \href {https://doi.org/10.1101/2023.04.28.538739}
  {\path{doi:10.1101/2023.04.28.538739}}.

\bibitem{Vitalis2009}
A.~Vitalis, R.~V. Pappu, {{ABSINTH}}: {{A}} new continuum solvation model for
  simulations of polypeptides in aqueous solutions, Journal of Computational
  Chemistry 30~(5) (2009) 673--699.
\newblock \href {https://doi.org/10.1002/jcc.21005}
  {\path{doi:10.1002/jcc.21005}}.

\bibitem{Souza2021}
P.~C.~T. Souza, R.~Alessandri, J.~Barnoud, S.~Thallmair, I.~Faustino,
  F.~Gr{\"u}newald, I.~Patmanidis, H.~Abdizadeh, B.~M.~H. Bruininks, T.~A.
  Wassenaar, P.~C. Kroon, J.~Melcr, V.~Nieto, V.~Corradi, H.~M. Khan,
  J.~Doma{\'n}ski, M.~Javanainen, H.~{Martinez-Seara}, N.~Reuter, R.~B. Best,
  I.~Vattulainen, L.~Monticelli, X.~Periole, D.~P. Tieleman, A.~H. {de Vries},
  S.~J. Marrink, Martini 3: A general purpose force field for coarse-grained
  molecular dynamics, Nature Methods 18~(4) (2021) 382--388.
\newblock \href {https://doi.org/10.1038/s41592-021-01098-3}
  {\path{doi:10.1038/s41592-021-01098-3}}.

\bibitem{Dignon2018}
G.~L. Dignon, W.~Zheng, Y.~C. Kim, R.~B. Best, J.~Mittal, Sequence determinants
  of protein phase behavior from a coarse-grained model, PLOS Computational
  Biology 14~(1) (2018) e1005941.
\newblock \href {https://doi.org/10.1371/journal.pcbi.1005941}
  {\path{doi:10.1371/journal.pcbi.1005941}}.

\bibitem{Dannenhoffer-Lafage2021}
T.~{Dannenhoffer-Lafage}, R.~B. Best, A {{Data-Driven Hydrophobicity Scale}}
  for {{Predicting Liquid}}--{{Liquid Phase Separation}} of {{Proteins}}, J.
  Phys. Chem. B 125~(16) (2021) 4046--4056.
\newblock \href {https://doi.org/10.1021/acs.jpcb.0c11479}
  {\path{doi:10.1021/acs.jpcb.0c11479}}.

\bibitem{Regy2021}
R.~M. Regy, J.~Thompson, Y.~C. Kim, J.~Mittal, Improved coarse-grained model
  for studying sequence dependent phase separation of disordered proteins,
  Protein Science 30~(7) (2021) 1371--1379.
\newblock \href {https://doi.org/10.1002/pro.4094}
  {\path{doi:10.1002/pro.4094}}.

\bibitem{Joseph2021}
J.~A. Joseph, A.~Reinhardt, A.~Aguirre, P.~Y. Chew, K.~O. Russell, J.~R.
  Espinosa, A.~Garaizar, R.~{Collepardo-Guevara}, Physics-driven coarse-grained
  model for biomolecular phase separation with near-quantitative accuracy, Nat
  Comput Sci 1~(11) (2021) 732--743.
\newblock \href {https://doi.org/10.1038/s43588-021-00155-3}
  {\path{doi:10.1038/s43588-021-00155-3}}.

\bibitem{Tesei2021}
G.~Tesei, T.~K. Schulze, R.~Crehuet, K.~{Lindorff-Larsen}, Accurate model of
  liquid--liquid phase behavior of intrinsically disordered proteins from
  optimization of single-chain properties, Proceedings of the National Academy
  of Sciences 118~(44) (2021) e2111696118.
\newblock \href {https://doi.org/10.1073/pnas.2111696118}
  {\path{doi:10.1073/pnas.2111696118}}.

\bibitem{Jussupow2025}
A.~Jussupow, D.~Bartley, L.~J. Lapidus, M.~Feig, {{COCOMO2}}: {{A
  Coarse-Grained Model}} for {{Interacting Folded}} and {{Disordered
  Proteins}}, Journal of Chemical Theory and Computation (Feb. 2025).
\newblock \href {https://doi.org/10.1021/acs.jctc.4c01460}
  {\path{doi:10.1021/acs.jctc.4c01460}}.

\bibitem{Lotthammer2024}
J.~M. Lotthammer, G.~M. Ginell, D.~Griffith, R.~J. Emenecker, A.~S. Holehouse,
  Direct prediction of intrinsically disordered protein conformational
  properties from sequence, Nature Methods 21~(3) (2024) 465--476.
\newblock \href {https://doi.org/10.1038/s41592-023-02159-5}
  {\path{doi:10.1038/s41592-023-02159-5}}.

\bibitem{sherry2017control}
K.~P. Sherry, R.~K. Das, R.~V. Pappu, D.~Barrick, Control of transcriptional
  activity by design of charge patterning in the intrinsically disordered {RAM}
  region of the {N}otch receptor, Proceedings of the National Academy of
  Sciences 114~(44) (2017) E9243--E9252.
\newblock \href {https://doi.org/10.1073/pnas.1706083114}
  {\path{doi:10.1073/pnas.1706083114}}.

\bibitem{Harmon2016}
T.~S. Harmon, M.~D. Crabtree, S.~L. Shammas, A.~E. Posey, J.~Clarke, R.~V.
  Pappu, {{GADIS}}: {{Algorithm}} for designing sequences to achieve target
  secondary structure profiles of intrinsically disordered proteins, Protein
  Engineering, Design and Selection 29~(9) (2016) 339--346.
\newblock \href {https://doi.org/10.1093/protein/gzw034}
  {\path{doi:10.1093/protein/gzw034}}.

\bibitem{Zeng2021}
X.~Zeng, C.~Liu, M.~J. Fossat, P.~Ren, A.~Chilkoti, R.~V. Pappu, Design of
  intrinsically disordered proteins that undergo phase transitions with lower
  critical solution temperatures, APL Materials 9~(2) (2021) 021119.
\newblock \href {https://doi.org/10.1063/5.0037438}
  {\path{doi:10.1063/5.0037438}}.

\bibitem{Pesce2024}
F.~Pesce, A.~Bremer, G.~Tesei, J.~B. Hopkins, C.~R. Grace, T.~Mittag,
  K.~{Lindorff-Larsen}, Design of intrinsically disordered protein variants
  with diverse structural properties, Science Advances 10~(35) (2024) eadm9926.
\newblock \href {https://doi.org/10.1126/sciadv.adm9926}
  {\path{doi:10.1126/sciadv.adm9926}}.

\bibitem{Krueger2024}
R.~Krueger, M.~P. Brenner, K.~Shrinivas, Generalized design of
  sequence-ensemble-function relationships for intrinsically disordered
  proteins, bioRxiv (2024).
\newblock \href {https://doi.org/10.1101/2024.10.10.617695}
  {\path{doi:10.1101/2024.10.10.617695}}.

\bibitem{Mollaei2024}
P.~Mollaei, D.~Sadasivam, C.~Guntuboina, A.~Barati~Farimani, {{IDP-Bert}}:
  {{Predicting Properties}} of {{Intrinsically Disordered Proteins Using Large
  Language Models}}, Journal of Physical Chemistry B 128~(49) (2024)
  12030--12037.
\newblock \href {https://doi.org/10.1021/acs.jpcb.4c02507}
  {\path{doi:10.1021/acs.jpcb.4c02507}}.

\bibitem{janson2023direct}
G.~Janson, G.~Valdes-Garcia, L.~Heo, M.~Feig, Direct generation of protein
  conformational ensembles via machine learning, Nature Communications 14~(1)
  (2023) 774.
\newblock \href {https://doi.org/10.1038/s41467-023-36443-x}
  {\path{doi:10.1038/s41467-023-36443-x}}.

\bibitem{Novak2025}
B.~Novak, J.~M. Lotthammer, R.~J. Emenecker, A.~S. Holehouse, Accurate
  predictions of conformational ensembles of disordered proteins with starling,
  bioRxiv (2025).
\newblock \href {https://doi.org/10.1101/2025.02.14.638373}
  {\path{doi:10.1101/2025.02.14.638373}}.

\bibitem{Janson2024}
G.~Janson, M.~Feig, Transferable deep generative modeling of intrinsically
  disordered protein conformations, PLOS Computational Biology 20~(5) (2024)
  e1012144.
\newblock \href {https://doi.org/10.1371/journal.pcbi.1012144}
  {\path{doi:10.1371/journal.pcbi.1012144}}.

\bibitem{Zhu2024}
J.~Zhu, Z.~Li, Z.~Zheng, B.~Zhang, B.~Zhong, J.~Bai, X.~Hong, T.~Wang, T.~Wei,
  J.~Yang, H.-F. Chen, Precise {{Generation}} of {{Conformational Ensembles}}
  for {{Intrinsically Disordered Proteins}} via {{Fine-tuned Diffusion
  Models}}, bioRxiv (2024).
\newblock \href {https://doi.org/10.1101/2024.05.05.592611}
  {\path{doi:10.1101/2024.05.05.592611}}.

\bibitem{zhang2025}
O.~Zhang, Z.~H. Liu, J.~D. Forman-Kay, T.~Head-Gordon, Deep learning of
  proteins with local and global regions of disorder (2025).
\newblock \href {https://doi.org/10.48550/arXiv.2502.11326}
  {\path{doi:10.48550/arXiv.2502.11326}}.

\bibitem{Schnapka2025}
V.~Schnapka, T.~I. Morozova, S.~Sen, M.~Bonomi, Atomic resolution ensembles of
  intrinsically disordered proteins with alphafold, bioRxiv (2025).
\newblock \href {https://doi.org/10.1101/2025.06.18.660298}
  {\path{doi:10.1101/2025.06.18.660298}}.

\bibitem{Lewis2025}
S.~Lewis, T.~Hempel, J.~Jiménez-Luna, M.~Gastegger, Y.~Xie, A.~Y.~K. Foong,
  V.~G. Satorras, O.~Abdin, B.~S. Veeling, I.~Zaporozhets, Y.~Chen, S.~Yang,
  A.~E. Foster, A.~Schneuing, J.~Nigam, F.~Barbero, V.~Stimper, A.~Campbell,
  J.~Yim, M.~Lienen, Y.~Shi, S.~Zheng, H.~Schulz, U.~Munir, R.~Sordillo,
  R.~Tomioka, C.~Clementi, F.~Noé, Scalable emulation of protein equilibrium
  ensembles with generative deep learning, Science 389~(6761) (2025) eadv9817.
\newblock \href {https://doi.org/10.1126/science.adv9817}
  {\path{doi:10.1126/science.adv9817}}.

\bibitem{Alamdari2024}
S.~Alamdari, N.~Thakkar, R.~van~den Berg, N.~Tenenholtz, R.~Strome, A.~M.
  Moses, A.~X. Lu, N.~Fusi, A.~P. Amini, K.~K. Yang, Protein generation with
  evolutionary diffusion: Sequence is all you need, bioRxiv (Nov. 2024).
\newblock \href {https://doi.org/10.1101/2023.09.11.556673}
  {\path{doi:10.1101/2023.09.11.556673}}.

\bibitem{Janson2025temp}
G.~Janson, A.~Jussupow, M.~Feig, Deep generative modeling of
  temperature-dependent structural ensembles of proteins, bioRxiv (2025).
\newblock \href {https://doi.org/10.1101/2025.03.09.642148}
  {\path{doi:10.1101/2025.03.09.642148}}.

\bibitem{mittal2018sequence}
A.~Mittal, A.~S. Holehouse, M.~C. Cohan, R.~V. Pappu, Sequence-to-conformation
  relationships of disordered regions tethered to folded domains of proteins,
  Journal of Molecular Biology 430~(16) (2018) 2403--2421.
\newblock \href {https://doi.org/10.1016/j.jmb.2018.05.012}
  {\path{doi:10.1016/j.jmb.2018.05.012}}.

\bibitem{taneja2021folded}
I.~Taneja, A.~S. Holehouse, Folded domain charge properties influence the
  conformational behavior of disordered tails, Current Research in Structural
  Biology 3 (2021) 216--228.
\newblock \href {https://doi.org/10.1016/j.crstbi.2021.08.002}
  {\path{doi:10.1016/j.crstbi.2021.08.002}}.

\bibitem{pajkos2025afflecto}
M.~Pajkos, I.~Clerc, C.~Zanon, P.~Bernadó, J.~Cortés, {AFflecto}: A web
  server to generate conformational ensembles of flexible proteins from
  {AlphaFold} models, Journal of Molecular Biology 437~(15) (2025) 169003.
\newblock \href {https://doi.org/10.1016/j.jmb.2025.169003}
  {\path{doi:10.1016/j.jmb.2025.169003}}.

\bibitem{mcbride2025predicting}
A.~C. McBride, F.~Yu, E.~H. Cheng, A.~Mpouli, A.~C. Soe, M.~Hammel, G.~T.
  Montelione, T.~G. Oas, S.~E. Tsutakawa, B.~R. Donald, Predicting pose
  distribution of protein domains connected by flexible linkers is an unsolved
  problem, bioRxiv (2025).
\newblock \href {https://doi.org/10.1101/2025.04.27.650885}
  {\path{doi:10.1101/2025.04.27.650885}}.

\bibitem{liu2025diffusing}
C.~Liu, K.~Wu, H.~Choi, H.~L. Han, X.~Zhang, J.~L. Watson, G.~Ahn, J.~Z. Zhang,
  S.~Shijo, L.~L. Good, et~al., Diffusing protein binders to intrinsically
  disordered proteins, Nature 644~(8077) (2025) 809–817.
\newblock \href {https://doi.org/10.1038/s41586-025-09248-9}
  {\path{doi:10.1038/s41586-025-09248-9}}.

\bibitem{Bhat2025}
S.~Bhat, K.~Palepu, L.~Hong, J.~Mao, T.~Ye, R.~Iyer, L.~Zhao, T.~Chen,
  S.~Vincoff, R.~Watson, T.~Z. Wang, D.~Srijay, V.~S. Kavirayuni, K.~Kholina,
  S.~Goel, P.~Vure, A.~J. Deshpande, S.~H. Soderling, M.~P. DeLisa,
  P.~Chatterjee, De novo design of peptide binders to conformationally diverse
  targets with contrastive language modeling, Science Advances 11~(4) (2025)
  eadr8638.
\newblock \href {https://doi.org/10.1126/sciadv.adr8638}
  {\path{doi:10.1126/sciadv.adr8638}}.

\bibitem{Chen2025PepMLM}
L.~T. Chen, Z.~Quinn, M.~Dumas, C.~Peng, L.~Hong, M.~Lopez-Gonzalez, A.~Mestre,
  R.~Watson, S.~Vincoff, L.~Zhao, J.~Wu, A.~Stavrand, M.~Schaepers-Cheu, T.~Z.
  Wang, D.~Srijay, C.~Monticello, P.~Vure, R.~Pulugurta, S.~Pertsemlidis,
  K.~Kholina, S.~Goel, M.~P. DeLisa, J.-T.~A. Chi, R.~Truant, H.~C. Aguilar,
  P.~Chatterjee, Target sequence-conditioned design of peptide binders using
  masked language modeling, Nature Biotechnology (2025).
\newblock \href {https://doi.org/10.1038/s41587-025-02761-2}
  {\path{doi:10.1038/s41587-025-02761-2}}.

\bibitem{vazquez2024novo}
S.~V{\'a}zquez~Torres, P.~J. Leung, P.~Venkatesh, I.~D. Lutz, F.~Hink, H.-H.
  Huynh, J.~Becker, A.~H.-W. Yeh, D.~Juergens, N.~R. Bennett, et~al., De novo
  design of high-affinity binders of bioactive helical peptides, Nature
  626~(7998) (2024) 435--442.
\newblock \href {https://doi.org/10.1038/s41586-023-06953-1}
  {\path{doi:10.1038/s41586-023-06953-1}}.

\bibitem{sahtoe2024design}
D.~D. Sahtoe, E.~A. Andrzejewska, H.~L. Han, E.~Rennella, M.~M. Schneider,
  G.~Meisl, M.~Ahlrichs, J.~Decarreau, H.~Nguyen, A.~Kang, et~al., Design of
  amyloidogenic peptide traps, Nature Chemical Biology 20~(8) (2024) 981--990.
\newblock \href {https://doi.org/10.1038/s41589-024-01578-5}
  {\path{doi:10.1038/s41589-024-01578-5}}.

\bibitem{ginell2025sequence}
G.~M. Ginell, R.~J. Emenecker, J.~M. Lotthammer, A.~T. Keeley, S.~P.
  Plassmeyer, N.~Razo, E.~T. Usher, J.~F. Pelham, A.~S. Holehouse,
  Sequence-based prediction of intermolecular interactions driven by disordered
  regions, Science 388~(6749) (2025) eadq8381.
\newblock \href {https://doi.org/10.1126/science.adq8381}
  {\path{doi:10.1126/science.adq8381}}.

\bibitem{dai2023programmable}
Y.~Dai, M.~Farag, D.~Lee, X.~Zeng, K.~Kim, H.-i. Son, X.~Guo, J.~Su,
  N.~Peterson, J.~Mohammed, et~al., Programmable synthetic biomolecular
  condensates for cellular control, Nature Chemical Biology 19~(4) (2023)
  518--528.
\newblock \href {https://doi.org/10.1038/s41589-022-01252-8}
  {\path{doi:10.1038/s41589-022-01252-8}}.

\bibitem{gil2024local}
M.~Gil-Garcia, A.~I. Ben{\'\i}tez-Mateos, M.~Papp, F.~Stoffel, C.~Morelli,
  K.~Normak, K.~Makasewicz, L.~Faltova, F.~Paradisi, P.~Arosio, Local
  environment in biomolecular condensates modulates enzymatic activity across
  length scales, Nature Communications 15~(1) (2024) 3322.
\newblock \href {https://doi.org/10.1038/s41467-024-47435-w}
  {\path{doi:10.1038/s41467-024-47435-w}}.

\bibitem{liang2024situ}
T.~Liang, Y.~Dong, I.~Cheng, P.~Wen, F.~Li, F.~Liu, Q.~Wu, E.~Ren, P.~Liu,
  H.~Li, et~al., In situ formation of biomolecular condensates as intracellular
  drug reservoirs for augmenting chemotherapy, Nature Biomedical Engineering
  8~(11) (2024) 1469--1482.
\newblock \href {https://doi.org/10.1038/s41551-024-01254-y}
  {\path{doi:10.1038/s41551-024-01254-y}}.

\bibitem{urry1974synthetic}
D.~Urry, M.~Long, B.~Cox, T.~Ohnishi, L.~Mitchell, M.~Jacobs, The synthetic
  polypentapeptide of elastin coacervates and forms filamentous aggregates,
  Biochimica et Biophysica Acta (BBA)-Protein Structure 371~(2) (1974)
  597--602.
\newblock \href {https://doi.org/10.1016/0005-2795(74)90057-9}
  {\path{doi:10.1016/0005-2795(74)90057-9}}.

\bibitem{urry1992hydrophobicity}
D.~W. Urry, D.~C. Gowda, T.~M. Parker, C.-H. Luan, M.~C. Reid, C.~M. Harris,
  A.~Pattanaik, R.~D. Harris, Hydrophobicity scale for proteins based on
  inverse temperature transitions, Biopolymers: Original Research on
  Biomolecules 32~(9) (1992) 1243--1250.
\newblock \href {https://doi.org/10.1002/bip.360320913}
  {\path{doi:10.1002/bip.360320913}}.

\bibitem{mcdaniel2013unified}
J.~R. McDaniel, D.~C. Radford, A.~Chilkoti, A unified model for de novo design
  of elastin-like polypeptides with tunable inverse transition temperatures,
  Biomacromolecules 14~(8) (2013) 2866--2872.
\newblock \href {https://doi.org/10.1021/bm4007166}
  {\path{doi:10.1021/bm4007166}}.

\bibitem{quiroz2015sequence}
F.~G. Quiroz, A.~Chilkoti, Sequence heuristics to encode phase behaviour in
  intrinsically disordered protein polymers, Nature Materials 14~(11) (2015)
  1164--1171.
\newblock \href {https://doi.org/10.1038/nmat4418}
  {\path{doi:10.1038/nmat4418}}.

\bibitem{dzuricky2020novo}
M.~Dzuricky, B.~A. Rogers, A.~Shahid, P.~S. Cremer, A.~Chilkoti, De novo
  engineering of intracellular condensates using artificial disordered
  proteins, Nature Chemistry 12~(9) (2020) 814--825.
\newblock \href {https://doi.org/10.1038/s41557-020-0511-7}
  {\path{doi:10.1038/s41557-020-0511-7}}.

\bibitem{Bremer2022}
A.~Bremer, M.~Farag, W.~M. Borcherds, I.~Peran, E.~W. Martin, R.~V. Pappu,
  T.~Mittag, Deciphering how naturally occurring sequence features impact the
  phase behaviours of disordered prion-like domains, Nature Chemistry 14~(2)
  (2022) 196--207.
\newblock \href {https://doi.org/10.1038/s41557-021-00840-w}
  {\path{doi:10.1038/s41557-021-00840-w}}.

\bibitem{Norrild2024}
R.~K. Norrild, S.~von B{\"u}low, E.~Halld{\'o}rsson, K.~Lindorff-Larsen, J.~M.
  Rogers, A.~K. Buell, Proteome-scale quantification of the interactions
  driving condensate formation of intrinsically disordered proteins, bioRxiv
  (2025).
\newblock \href {https://doi.org/10.1101/2024.12.21.629870}
  {\path{doi:10.1101/2024.12.21.629870}}.

\bibitem{Farag2022}
M.~Farag, S.~R. Cohen, W.~M. Borcherds, A.~Bremer, T.~Mittag, R.~V. Pappu,
  Condensates formed by prion-like low-complexity domains have small-world
  network structures and interfaces defined by expanded conformations, Nature
  Communications 13~(1) (2022) 7722.
\newblock \href {https://doi.org/10.1038/s41467-022-35370-7}
  {\path{doi:10.1038/s41467-022-35370-7}}.

\bibitem{lichtinger2021targeted}
S.~M. Lichtinger, A.~Garaizar, R.~Collepardo-Guevara, A.~Reinhardt, Targeted
  modulation of protein liquid--liquid phase separation by evolution of
  amino-acid sequence, PLOS Computational Biology 17~(8) (2021) e1009328.
\newblock \href {https://doi.org/10.1371/journal.pcbi.1009328}
  {\path{doi:10.1371/journal.pcbi.1009328}}.

\bibitem{chew2023thermodynamic}
P.~Y. Chew, J.~A. Joseph, R.~Collepardo-Guevara, A.~Reinhardt, Thermodynamic
  origins of two-component multiphase condensates of proteins, Chemical Science
  14~(7) (2023) 1820--1836.
\newblock \href {https://doi.org/10.1039/D2SC05873A}
  {\path{doi:10.1039/D2SC05873A}}.

\bibitem{jin2025predicting}
J.~Jin, W.~Oliver, M.~A. Webb, W.~M. Jacobs, Predicting heteropolymer phase
  separation using two-chain contact maps, The Journal of Chemical Physics
  163~(1) (2025).
\newblock \href {https://doi.org/10.1063/5.0269504}
  {\path{doi:10.1063/5.0269504}}.

\bibitem{oliver2025b2enoughevaluatingsimple}
W.~W. Oliver, W.~M. Jacobs, M.~A. Webb, When ${B}_2$ is not enough: Evaluating
  simple metrics for predicting phase separation of intrinsically disordered
  proteins (2025).
\newblock \href {https://doi.org/10.48550/arXiv.2507.12312}
  {\path{doi:10.48550/arXiv.2507.12312}}.

\bibitem{vonBulow2025}
S.~von Bülow, G.~Tesei, F.~K. Zaidi, T.~Mittag, K.~Lindorff-Larsen, Prediction
  of phase-separation propensities of disordered proteins from sequence,
  Proceedings of the National Academy of Sciences 122~(13) (2025) e2417920122.
\newblock \href {https://doi.org/10.1073/pnas.2417920122}
  {\path{doi:10.1073/pnas.2417920122}}.

\bibitem{An2024}
Y.~An, M.~A. Webb, W.~M. Jacobs, Active learning of the thermodynamics-dynamics
  trade-off in protein condensates, Science Advances 10~(1) (2024) eadj2448.
\newblock \href {https://doi.org/10.1126/sciadv.adj2448}
  {\path{doi:10.1126/sciadv.adj2448}}.

\bibitem{Changiarath2025}
A.~Changiarath, A.~Arya, V.~A. Xenidis, J.~Padeken, L.~S. Stelzl, Sequence
  determinants of protein phase separation and recognition by protein
  phase-separated condensates through molecular dynamics and active learning,
  Faraday Discussions 256 (2025) 235--254.
\newblock \href {https://doi.org/10.1039/d4fd00099d}
  {\path{doi:10.1039/d4fd00099d}}.

\bibitem{Kilgore2025}
H.~R. Kilgore, I.~Chinn, P.~G. Mikhael, I.~Mitnikov, C.~Van~Dongen,
  G.~Zylberberg, L.~Afeyan, S.~F. Banani, S.~Wilson-Hawken, T.~I. Lee,
  R.~Barzilay, R.~A. Young, Protein codes promote selective subcellular
  compartmentalization, Science 387~(6738) (2025) 1095--1101.
\newblock \href {https://doi.org/10.1126/science.adq2634}
  {\path{doi:10.1126/science.adq2634}}.

\bibitem{Ausserwoeger2024}
H.~Ausserw{\"o}ger, R.~Scrutton, C.~M. Fischer, T.~Sneideris, D.~Qian,
  E.~de~Csill{\'e}ry, I.~Baronaite, K.~L. Saar, A.~Z. Bia{\l}ek, M.~Oeller,
  G.~Krainer, T.~M. Franzmann, S.~Wittmann, J.~M. Iglesias-Artola,
  G.~Invernizzi, A.~A. Hyman, S.~Alberti, N.~Lorenzen, T.~P.~J. Knowles,
  Biomolecular condensates sustain ph gradients at equilibrium through charge
  neutralisation, bioRxiv (2025).
\newblock \href {https://doi.org/10.1101/2024.05.23.595321}
  {\path{doi:10.1101/2024.05.23.595321}}.

\bibitem{posey2024biomolecular}
A.~E. Posey, A.~Bremer, N.~A. Erkamp, A.~Pant, T.~P. Knowles, Y.~Dai,
  T.~Mittag, R.~V. Pappu, Biomolecular condensates are characterized by
  interphase electric potentials, Journal of the American Chemical Society
  146~(41) (2024) 28268--28281.
\newblock \href {https://doi.org/10.1021/jacs.4c08946}
  {\path{doi:10.1021/jacs.4c08946}}.

\bibitem{Linsenmeier2023}
M.~Linsenmeier, L.~Faltova, C.~Morelli, U.~Capasso~Palmiero, C.~Seiffert, A.~M.
  K\"{u}ffner, D.~Pinotsi, J.~Zhou, R.~Mezzenga, P.~Arosio, The interface of
  condensates of the {hnRNPA1} low-complexity domain promotes formation of
  amyloid fibrils, Nature Chemistry 15~(10) (2023) 1340–1349.
\newblock \href {https://doi.org/10.1038/s41557-023-01289-9}
  {\path{doi:10.1038/s41557-023-01289-9}}.

\bibitem{das2025tunable}
T.~Das, F.~K. Zaidi, M.~Farag, K.~M. Ruff, T.~S. Mahendran, A.~Singh, X.~Gui,
  J.~Messing, J.~P. Taylor, P.~R. Banerjee, et~al., Tunable metastability of
  condensates reconciles their dual roles in amyloid fibril formation,
  Molecular Cell 85~(11) (2025) 2230--2245.
\newblock \href {https://doi.org/10.1016/j.molcel.2025.05.011}
  {\path{doi:10.1016/j.molcel.2025.05.011}}.

\bibitem{visser2024controlling}
B.~S. Visser, M.~H. van Haren, W.~P. Lipi{\'n}ski, K.~A. van
  Leijenhorst-Groener, M.~M. Claessens, M.~V. Queir{\'o}s, C.~H. Ramos,
  J.~Eeftens, E.~Spruijt, Controlling interfacial protein adsorption,
  desorption and aggregation in biomolecular condensates, bioRxiv (2024).
\newblock \href {https://doi.org/10.1101/2024.10.20.619145}
  {\path{doi:10.1101/2024.10.20.619145}}.

\bibitem{vanHilten2023}
N.~{van Hilten}, J.~Methorst, N.~Verwei, H.~J. Risselada, Physics-based
  generative model of curvature sensing peptides; distinguishing sensors from
  binders, Science Advances 9~(11) (2023) eade8839.
\newblock \href {https://doi.org/10.1126/sciadv.ade8839}
  {\path{doi:10.1126/sciadv.ade8839}}.

\bibitem{Janson2025}
G.~Janson, M.~Feig, Generation of protein dynamics by machine learning, Current
  Opinion in Structural Biology 93 (2025) 103115.
\newblock \href {https://doi.org/10.1016/j.sbi.2025.103115}
  {\path{doi:10.1016/j.sbi.2025.103115}}.

\end{thebibliography}

\end{document}